# Are there Helium-like Protonic States of Individual Water Molecules in Liquid H₂O?

Ulrich Müller-Herold[1]

*ETH Zurich, P.O. Box CAB 42, CH-8092 Zurich, Switzerland*

Dedicated to the memory of Prof. Anton Amann (1956 – 2015)

***Abstract*** Are there indications that individual $H_2O$ molecules in liquid water can loose their bent structure, i.e. that the protons give up their rigid angular correlation and behave largely uncorrelated, similar to electrons in the ground-state of helium? In agreement with the two-state picture of liquid water this would allow for the thermal coexistence of tetraedrically coordinated and spherical water molecules in the liquid.

In the Hooke-Calogero model of a confined triatomic of $XY_2$-type it is shown that energetically low-lying zero orbital-momentum states, which are bent if unconfined can change to helium-like shape under increasing confinement strength $f$. For the respective states this occurs at different values for $f$. It turns out that at $f = 2.79$ a bent and a helium-like state can thermally coexist. In order to characterize more precisely 'helium-like' angular correlation a maximum entropy estimate for the marginal correlation of electrons in the helium ground state is given.

---
[1] e-mail: mueller-herold@env.ethz.ch

# 1 Introduction

Some time ago it has been argued that certain properties of normal and heavy water are similar to those of noble-gas like simple liquids: An analysis in terms of corresponding states suggests that the temperature dependence of the specific volume per molecule and the heat of evaporation on the coexistence curve of water show an argon-like behavior [1]. The same applies to the kinematic shear viscosity [2]. These observations lead to more general questions about possible relations between the water molecule and particular noble gases. At first sight, Neon is the most obvious candidate: $H_2O$ and Ne are not only iso-electronic but, moreover, Ne is the united atom arising if the protons and the oxygen nucleous coalesce. Since long it is known that this has far-reaching conse-quences for Neon ground-state orbitals and their relation to the electronic structure of $H_2O$ [3].

The clue to still more fundamental parallels results from the hybrid nature of the protons being the lightests of all nuclei. On the one side they are treated on equal (quasiclassical) footing with the heavier nuclei in the construction of potential energy surfaces in molecular quantum mechnics [4]. On the other they can behave like fully quantum-mechanical lighter particles, e.g. like electrons or myons. This makes it temp-ting to look for states where $H_2O$ looses the quasi-rigid bent structure known from isolated and from tetraedrically hydrogen-bridged water molecules. The most pronounced conceivable quantum-mechanical comportment would be a state in which the two protons largely lose their angular correlation and behave similarly to electrons in the ground state of helium (with $O^{2-}$ playing a role analogous to $He^{2+}$). At first sight this may be in contradiction to long-standing experimental and theoretical knowledge on crystalline ice and low-pressure water vapour. However in connection with the findings on amorphous ice and supercooled liquid water [5] which are interpreted  as an interplay of  tetraedrically structured low-density molecules and „unstructured" closest-packed high-density molecules the tetraedral picture somewhat looses its implicitness.

In an earlier publication [6] it was demonstrated in a Hooke-Calogero model that confined three-particle systems of $XY_2$-type can change from a bent, directed-bond like structure to helium-like angular correlation of the two equal particles if the confinement strength is increased. In order to link this to the water problem, one has to show that the energetically lowest lying eigenstates show different behavior under confinement: This would allow for the thermal coexistence of individual water molecules of different type in the presence of liquid water as a confining macroscopic substance.

In the present paper we investigate the behavior of the six lowest-lying zero orbital momentum eigenstates of the Hooke-Calogero model under varying confinement strenght. It turns out that there are two states which show the mentioned behavior: Without confinement they are V-shaped („bent"). At a well-defined confinement strength one of the states changes to helium-like $(1s)^2$-type behavior of the lighter Y-particles („protons")  whereas the other one essentially keeps its bent shape. Moreover, the ideal type of „helium-like angular correlation" is characterized through a maximum entropy estimate of marginal radial correlation in the ground state of helium.

## 2 The isolated Hooke-Calogero system

The Hooke-Calogero Hamiltonian $\hat{H}_{mol}$ [6] is a modification of the harmonium or Hookium three-body Hamiltionian where all attracting forces are harmonic and repulsion is given by an $1/r^2$ potential

$$\hat{H}_{mol} = -\frac{1}{2m_e}\Delta_{\vec{Q}_1} - \frac{1}{2m_e}\Delta_{\vec{Q}_2} - \frac{1}{2m_u}\Delta_{\vec{Q}_3} + V \qquad (1)$$

$$V = \frac{1}{2}\left(\vec{Q}_3 - \vec{Q}_1\right)^2 + \frac{1}{2}\left(\vec{Q}_3 - \vec{Q}_2\right)^2 + \frac{1}{\left(\vec{Q}_2 - \vec{Q}_1\right)^2} + \frac{15}{4}\left(\vec{Q}_2 - \vec{Q}_1\right)^2 \qquad (2)$$

$\vec{Q}_1, \vec{Q}_2$ and $m_e$ are the coordinates and the mass of the two equal Y particles, whereas $\vec{Q}_3$ and $m_u$ are the coordinates and the mass of the third "unequal" X particle. As in atomic units Planck's constant is unity. The Y mass is set to $m_e = 2$ and the X mass is assumed to be $m_u = 32$. (Accordingly, as in the water molecule, the mass ratio between the unequal particle and the equal ones is 16.) In Jacobi coordinates

$$\begin{pmatrix} \vec{r}_1 \\ \vec{r}_2 \\ \vec{r}_3 \end{pmatrix} = \begin{pmatrix} -1 & 1 & 0 \\ -1/2 & -1/2 & 1 \\ m_e/M & m_e/M & m_u/M \end{pmatrix} \begin{pmatrix} \vec{Q}_1 \\ \vec{Q}_2 \\ \vec{Q}_3 \end{pmatrix} \tag{3}$$

the Hamiltonian reads

$$\hat{H}_{mol} = -\frac{1}{m_e}\Delta_{\vec{r}_1} - \frac{1}{2\mu}\Delta_{\vec{r}_2} - \frac{1}{2M}\Delta_{\vec{r}_3} + 4\vec{r}_1^{\,2} + \frac{1}{\vec{r}_1^{\,2}} + \vec{r}_2^{\,2} \tag{4}$$

where $M = 2\,m_e + m_u = 36$ denotes the total mass and $\mu = 2\,m_e\,m_u/(2\,m_e + m_u) = 32/9$ is the reduced mass of the three-body system. $\vec{r}_3$ denotes the center-of-mass coordinate.

## 3 Confinement

„Quite generally one needs to make a distinction between a hypothetical isolated molecule, and a really observed *individual* molecule. If a molecule is not isolated it must be interacting with something; that something is loosely referred to as the 'environment'. It might be other molecules, the (macroscopic) substance the molecule finds itself in, or quantized electromagnetic radiation. The interesting question is how to get from the quantum theory of an isolated molecule to a quantum theory of an individual molecule by rational mathematics." [7]

The probably simplest way to model a molecule's environment is via confining harmonic potentials such as e.g. in the theory of quantum dots [6]. In contrast to solids, confinement in liquids does not localize the object under consideration in a fixed spatial domain but in a solvent „cage" moving together with the confined object. Accordingly, the confining potential does not act on the object's center-of-mass coordinate, but only on the relative coordinates in the center-of-mass system. As in ref. [6] the confinement potential is

$$\begin{aligned} V_{conf} &= \frac{f'}{2}\left\{ m_e^{\,2}\left(\vec{Q}_1 - \vec{r}_3\right)^2 + m_e^{\,2}\left(\vec{Q}_2 - \vec{r}_3\right)^2 + m_u^{\,2}\left(\vec{Q}_3 - \vec{r}_3\right)^2 \right\} \\ &= \frac{f}{2}\left\{ \frac{\vec{r}_1^{\,2}}{2} + \frac{6m_u^{\,2}}{M^2}\vec{r}_2^{\,2} \right\} \end{aligned} \tag{5}$$

where $f := f' m_e^2$ denotes the confinement strength. In Jacobi coordinates the total Hamiltonian $\hat{H} = \hat{H}_{mol} + \hat{V}_{conf}$ decouples into three commuting parts $\hat{H} = \hat{H}_1 + \hat{H}_2 + \hat{H}_3$ (For further details see [6].) After substraction of the center of mass part $\hat{H}_3$ the internal Hamiltonian reads $\hat{H}_{int} := \hat{H} - \hat{H}_3 = \hat{H}_1 + \hat{H}_2$ with

$$\hat{H}_1 := -\frac{1}{2}\Delta_{\vec{r}_1} + 4\vec{r}_1^{\,2} + \frac{f}{4}\vec{r}_1^{\,2} + \frac{1}{\vec{r}_1^{\,2}} \tag{8}$$

$$\hat{H}_2 := -\frac{9}{64}\Delta_{\vec{r}_2} + \vec{r}_2^{\,2} + \frac{f}{2}\frac{128}{27}\vec{r}_2^{\,2} \tag{9}$$

## 4 Angular correlation

The total ground-state wave function of $\hat{H}_{int}$ is the product of the respective ground-state wave functions of Hamiltonians $\hat{H}_1$ and $\hat{H}_2$

$$\Psi_{1,0}(r_1) = r_1 e^{-a r_1^2/4} \tag{10}$$

$$\Psi_{2,0}(r_2) = e^{-b r_2^2} \tag{11}$$

with $a := 2\sqrt{8 + f/2}$, $b := \sqrt{(8/9)(2 + 128\,f/27)}$, and $r_i := |\vec{r}_i|$. Molecular shape is extracted from the total wave function along the lines of reference [8]: beginning with the ground-state density

$$\left(\Psi_{1,0}\Psi_{1,0}^*\right)(\vec{r}_1)\left(\Psi_{2,0}\Psi_{2,0}^*\right)(\vec{r}_2) = r_1^2 e^{-a r_1^2/2} e^{-2 b r_2^2} \tag{12}$$

one calculates the ground-state expectation value of the two-density operator

$$\begin{aligned}\hat{\rho}'(\vec{q}_1, \vec{q}_2) &= \delta(\vec{Q}_3 - \vec{Q}_1 - \vec{q}_1)\delta(\vec{Q}_3 - \vec{Q}_2 - \vec{q}_2) \\ &= \delta(\vec{r}_2 - \vec{r}_1/2 - \vec{q}_1)\delta(\vec{r}_2 + \vec{r}_1/2 - \vec{q}_2)\end{aligned} \tag{13}$$

which gives

$$\rho'_{00}(\vec{q}_1, \vec{q}_2) \propto |\vec{q}_1 - \vec{q}_2|^2 e^{-a(\vec{q}_1 - \vec{q}_2)^2/2} e^{-b(\vec{q}_1 + \vec{q}_2)^2/2} \tag{14}$$

The probability of finding the Y particles at a distance $q$ from the X particle in a shell of thickness d$q$ is proportional to

$$\mathrm{d}q\, q^2\, \rho'_{00}(\vec{q}_1, \vec{q}_2)_{q_1 = q_2 = q} = \mathrm{d}q\, q^4\, (2 - 2\cos\alpha)\, e^{-aq^2(1-\cos\alpha)}\, e^{-bq^2(1+\cos\alpha)} \tag{15}$$

with $\alpha$ being the Y-X-Y "bond" angle. Integration over $q$ now leads to the anglar ground-state probability distribution as a function of $\alpha$ alone

$$\rho_{00}(\alpha) \propto \int_0^\infty \mathrm{d}q\, q^2\, \rho'_{00}(\vec{q}_1, \vec{q}_2)_{q_1 = q_2 = q} = \frac{1 - \cos\alpha}{\left(a(1-\cos\alpha) + b(1+\cos\alpha)\right)^{5/2}} \tag{16}$$

which is depicted for the isolated (f = 0) and the confined (f = 2.79) case in FIG 1. For excited states of $\hat{H}_{\mathrm{int}}$ the angular probability distributions are obtained in the same way: if $j$ denotes the $j$-th excited state of $\hat{H}_1$ and $k$ the $k$-th excited state of $\hat{H}_2$ one begins with the density $\left(\Psi_{1,j}\Psi_{1,j}^*\right)(\vec{r}_1)\left(\Psi_{2,k}\Psi_{2,k}^*\right)(\vec{r}_2)$ as in eq. (12) and ends with $\rho_{jk}(\alpha)$. (Appendix A)

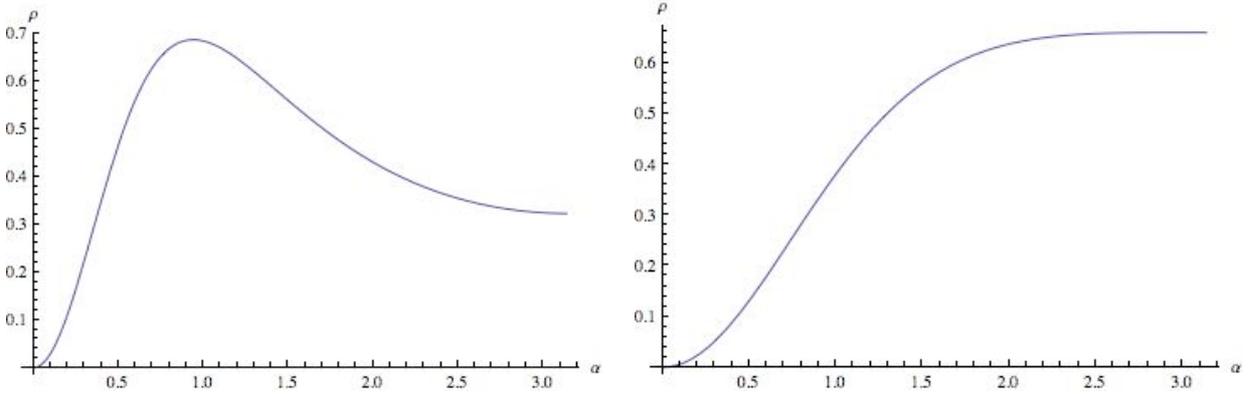

FIG. 1 The marginal ground-state probability distribution $\rho$ of the Y-X-Y angle for the isolated (f = 0, left) and the confined (f = 2.79, right) XY$_2$ system. The isolated XY$_2$ has a bent structure with a maximum of $\rho$ at $\alpha = 0.945$ corresponding to a „bond angle" of 54°. (From Ref. [6])

## 5 Ideal helium-like angular correlation

The nominal configuration of the ${}^1$S ground state of helium is $(1s)^2$. In such a state there is no radial or angular correlation and since the electrons have opposite spin there is no Fermi hole. In the approximative Hartree-Fock description a small amount of *radial* correlation is introduced because there is no angular correlation and the variationally lowered exponent of the helium 1s-orbitals pushes the electrons away from the nucleus and hence from each other [9]. The absence of angular correlation manifests itself in a constant probability distribution $\rho$ of the X-Y-X angle $\alpha$ with $\rho = \frac{1}{2}$ due to normalization.

A simple geometrical characterization of electron correlation in the *exact* helium ground state is difficult mainly for two reasons:

- There is almost too much spatial information in highly complex three-variable wave functions $\Psi(q_1, q_2, \alpha)$ [10]. Simple intuitive pictures arise only after averaging over irrelevant details. As for angular correlation the marginal distribution of the X-Y-X angle $\alpha$ is particularly useful since it eliminates the residual radial correlation. Going one step further R.S. Berry and various coauthors investigated two-electron states on spheres with fixed radius and Coulomb repulsion [11]. This rigid-bender type model [12] provides a schematic picture of intrashell angular correlation which is essentially reproduced by full three-variable densities if one integrates over $q_1$ and $q_2$ keeping $q_1 = q_2$.

- Furthermore it is an open question to what extent energetically nearly exact wave functions approximate other properties of the exact solution. (The energetically almost exact 26-term Hylleras-Kinoshita helium wave function - with ground-state energy of -2.903722 hartree (exact value: -2.903724) - does not even fulfill the cusp condition [13].)

For these reasons we choose a more direct approach and design the ideal type of angular correlation in the ground state of helium via the maximum entropy formalism. (See Appendix B) This leads to a minimally prejudiced probablity distribution $\rho_{ME}$ for $\alpha$. For finite expectation values of interelectronic Coulomb repulsion it is given by

$$\rho_{ME} = e^{-\mu/\sqrt{1-\cos\alpha}} / N \tag{17}$$

(FIG. 2) where $\mu = 0.6$ is a Lagrangian multiplier and N = 0.992 is the normalization constant.

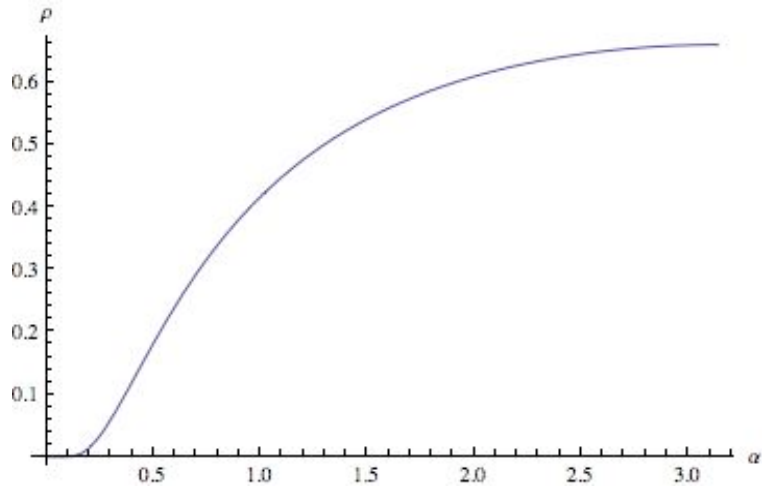

FIG. 2  Maximum entropy probabilty distribution $\rho_{\text{ME}}$ of electronic angular
correlation in the helium ground state (App. B)

We are now in a position to compare the angular correlation in the ground state of the
confined $XY_2$ molecule as depicted in FIG. 1 with the ideal helium-like angular correla-
tion obtained via the maximum entropy formalism (FIG. 2). In FIG. 3 one sees that the
two marginal probability distributions are quite similiar. This is confirmed by the small

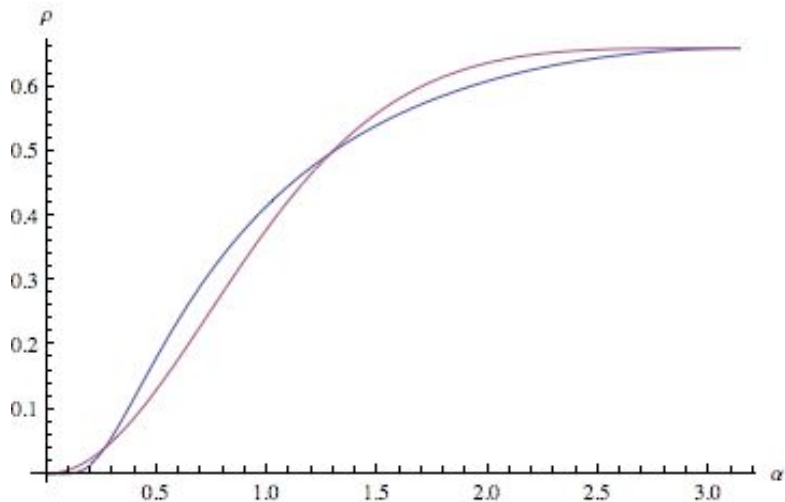

FIG. 3 Comparision of the maximum entropy angular probability distribution $\rho_{\text{ME}}$ with
the ground-state angular probability distribution $\rho_{00}$ of the confined $XY_2$ molecule.
The small Hellinger distance documents that they are almost identical.

value of their Hellinger distance $H(\rho_{ME}, \rho_{00}) = 0.027$ where

$$H^2(\rho_{ME}, \rho_{00}) \quad = \quad \frac{1}{2}\int_0^\pi \left(\sqrt{\rho_{ME}} - \sqrt{\rho_{00}}\right)^2 \sin\alpha \; d\alpha \quad = \quad 0.00075 \tag{18}$$

(The Hellinger distance between two probability distributions is zero if they are identical. The maximal distance 1 is achieved if the distributions are maximally different, i.e. if they are singular with respect to each other.)

## 6 Excited states under confinement

From the five lowest-lying excited states with zero angular-momentum and marginal angular probability distributions $\rho_{10}, \rho_{01}, \rho_{20}, \rho_{11}$ and $\rho_{02}$ resp. only $\rho_{20}$ (FIG. 4) shows a simple behavior under confinement, comparable to the ground-state (FIG. 1). The bent structure of the isolated molecule ($f$ = 0) is even more pronounced than in the ground state, the bond angle is 26.6°. Increasing confinement strength broadens the peak and shifts it toward higher values of $\alpha$. Although a second maximum develops at $\alpha = \pi$ the bent structure survives at maximal confinement ($f$ = 2.79) with „bond angle" $\alpha$ = 0.793 = 45.4°. (The unphysical value $f$ = -0.3 is included for mathematical completeness.)

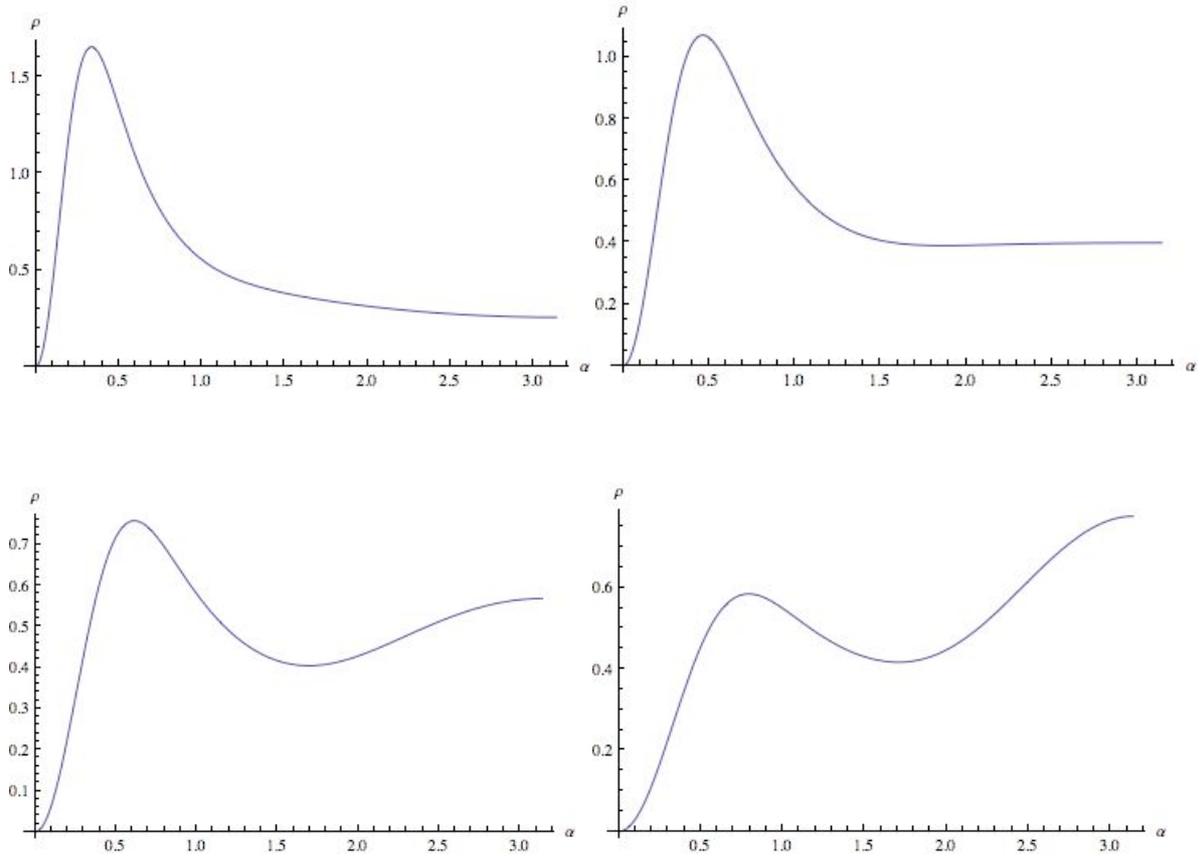

FIG. 4 Marginal angular probability distribution $\rho$ in the $\Psi_{1,2}\Psi_{2,0}$ state under varying confinement strength $f$: -0.3 upper left, 0 upper right, 0.79 lower left, 2.79 lower right with corresponding maxima at 0.336, 0.465, 0.614, and 0.793 (in radians).

Due to the behavior of $\rho_{20}$ the situation can be summarized as follows: In the modified Hooke-Calogero model of $XY_2$ [6] which stands as a proxy for the nuclear dynamics of the water molecule two different energetically low-lying molecular forms can coexist in the presence of liquid water as a confining macroscopic medium, a bent-one and a nobel-gas like spherical one. As in the isolated $H_2O$ molecule the corresponding states $\Psi_{1,0}\Psi_{2,0}$ and $\Psi_{1,2}\Psi_{2,0}$ of the isolated $XY_2$ are both bent.

## 7 Discussion

Till this day the two-component picture of liquid water has gained but partial acceptance. The main reason is probably the lack of a simple and intuitve molecular-level picture of the second component. Whereas the structure of the first component – the so-called *ice molecules* as they where named by W.C. Roentgen long ago [14] – can easily be traced back to frozen water in its Ih state there is no equally plausible candidate for the second constituent.

Traditionally, the ice-type molecules are associated with approximately tetraedrical bond angles, hydrogen bridges between neighboring oxygen nuclei und a local coordination of four. The experimental facts in favour of the two-component picture suggest that the second-constituent molecules are closest-packed [5] and have slightly higher molar energy and entropy [15].

In principle, quantum mechanics can be expected to answer the question. In practice, however, this is hampered by difficulties in connection with the use of two types of approximations which possibly mask the effect to be looked for: the Born-Oppenheimer or clamped-nuclei approximation and the Heisenberg cut, leading to a well-defined individual $H_2O$-(quasi)-particle in the presence of liquid water:

- the hybrid nature of the protons suggests that the clamped-nuclei approximation should be replaced with an all-particle wave function treating electrons and protons on the same footing. The stochastic variational optimization of explicitly correlated Gaussian geminals [16 ,17] lends itself to corresponding all-particle calculations *once a computationally manageable antisymmetrization procedure for the ten electrons of the water molecule has been found;*

- fundamentally, the discussion of an individual $H_2O$ (quasi)-particle within liquid water as a solvent implies a suitable decomposition of the total wave function into a tensor product of one function for the (quasi)-particle and one for the solvent, together with the specification of an effective Hamiltonian for the (quasi)-particle.

The choice of a suitable cut between a (quasi)-particle and its environment is by no means trivial. Among other it depends on how much of the environment is included in the (quasi)-particle Hamiltonian, which in the present paper is the sum of the isolated-particle Hamiltonian (eq. 4) plus a dressing part (eq. 5) taking confinement as the relevant effective influence of the environment. Of course, there are many other possibilities of modelling confinement [18] and there are other concepts of tensor-product decompositions than confinement such as e.g. the Hartree factorization [19].

In such a situation the analysis of simple solvable models is a viable way to adress questions where solutions for realistic models are out of reach. They can provide qualitative insight which often paves the way for the ultimate solution.

## Appendix A

In order to obtain $\rho_{20}(\alpha)$ from the non-normalized zero orbital-momentum eigen-functions of $\hat{H}_1$ and $\hat{H}_2$

$$\Psi_{1,0}(r_1) = r_1 e^{-a\, r_1^2/4} \qquad\qquad \Psi_{2,0}(r_2) = e^{-b\, r_2^2}$$

$$\Psi_{1,1}(r_1) = r_1 e^{-a\, r_1^2/4}(1 - \frac{2}{5}\frac{a}{2}r_1^2) \qquad \Psi_{2,1}(r_2) = e^{-b\, r_2^2}(1 - \frac{4}{3}b\, r_2^2)$$

$$\Psi_{1,2}(r_1) = r_1 e^{-a\, r_1^2/4}(1 - \frac{4}{5}\frac{a}{2}r_1^2 + \frac{4}{35}\frac{a^2}{2^2}r_1^4) \quad \Psi_{2,2}(r_2) = e^{-b\, r_2^2}(1 - \frac{8}{3}b\, r_2^2 + \frac{16}{15}b^2 r_2^4)$$

one first calculates the expectation value of the two-density operator $\hat{\rho}'$ (Eq. 13) with respect to $(\Psi_{1,2}\Psi_{1,2}^*)(\vec{r}_1)(\Psi_{2,0}\Psi_{2,0}^*)(\vec{r}_2)$. With $r_1^2 = (\vec{q}_1 - \vec{q}_2)^2$ and $r_2^2 = (\vec{q}_1 + \vec{q}_2)^2/4$ this gives

$$\rho'_{20}(\vec{q}_1, \vec{q}_2) \quad \propto \quad \rho'_{00}(\vec{q}_1, \vec{q}_2)\left(1 - \frac{4}{5}\frac{a}{2}(\vec{q}_1 - \vec{q}_2)^2 + \frac{4}{35}\frac{a^2}{2^2}(\vec{q}_1 - \vec{q}_2)^4\right)^2 \qquad (A1)$$

where $\rho'_{00}$ is taken from Eqs. 14 and 15. The probability of finding the Y particles at a distance $q$ from the X particle in a shell of thickness d$q$ is then

$$\mathrm{d}q\, q^2 \rho'_{20}(\vec{q}_1, \vec{q}_2)_{q_1=q_2=q} \propto \mathrm{d}q\, q^2 \left(1 - \frac{4}{5}a q^2(1-\cos\alpha) + \frac{4}{35}a^2 q^4(1-\cos\alpha)^2\right)^2$$
$$\times \rho'_{00}(\vec{q}_1, \vec{q}_2)_{q_1=q_2=q} \qquad (A2)$$

The integration leading to the marginal probability distribution of the bond angle

$$\rho_{20}(\alpha) \quad \propto \quad \int_0^\infty \mathrm{d}q\, q^2 \rho'_{20}(\vec{q}_1, \vec{q}_2)_{q_1=q_2=q} \qquad (A3)$$

involves several integrals of type

$$J_n := \int_0^\infty \mathrm{d}q\, q^{4+n} e^{-a q^2(1-\cos\alpha)} e^{-b q^2(1+\cos\alpha)}$$
$$= \frac{(3+n)!!\sqrt{\pi}}{8\, 2^{n/2}\big(a(1-\cos\alpha) + b(1+\cos\alpha)\big)^{(5+n)/2}} \qquad (A4)$$

with n even and $(3+n)!! = 1\cdot 3\cdot 5\cdot \ \cdot(3+n)$. It leads to the final result

$$\rho_{20}(\alpha) \;\propto\; J_0 - \frac{8}{5}\,a\,J_2(1-\cos\alpha) + \frac{152}{175}\,a^2\,J_4(1-\cos\alpha)^2 -$$
$$\frac{32}{175}\,a^3\,J_6(1-\cos\alpha)^3 + \frac{16\,a^4\,J_8(1-\cos\alpha)^4}{1225} \tag{A5}$$

The calculation of angular distributions $\rho_{jk}(\alpha)$, $(j,k)\neq(2,0)$ for other excited states proceeds along the same lines.

## Appendix B

The maximum entropy formalism infers minimally prejudiced probability distributions from empirical observations [20]. A probability distribution is minimally prejudiced if it is as random as possible, i.e. if has the maximal Shannon entropy of all probability distributions compatible with the given empirical facts. As for the ground state of helium we know that the behavior of the electrons is completely uncorrelated as long as the Coulomb repulsion is switched off. In addition we know that in the true ground state the expectation value of Coulomb repulsion has a unknown, but well-determined finite value. The maximum entropy formalism now gives the maximally uncorrelated marginal angular probability distribution $\rho_{\mathrm{ME}}$ given that the expectation of Coulomb repulsion has a finite value.

The internal zero orbital momentum wave functions of helium-like atoms can be written as a function $\Psi$ of three variables $q_1$, $q_2$ and $\alpha$ where the $q_i$ denote the distance of the two electrons from the nucleus and $\alpha$ is the interelectronic angle. If the Coulomb repulsion

$$\frac{1}{|\vec{q}_1 - \vec{q}_2|} \;=\; \frac{1}{(q_1^2 + q_2^2 - 2q_1 q_2 \cos\alpha)^{1/2}} \tag{B1}$$

is restricted to shells of radius $q = q_1 = q_2$ its expectation value reads

$$\int_0^\infty \mathrm{d}q \int_0^\pi \sin\alpha\,\mathrm{d}\alpha\, \frac{(\Psi\Psi^*)(q,q,\alpha)}{q\sqrt{2(1-\cos\alpha)}} \;=\; \int_0^\pi \frac{\rho(\alpha)}{\sqrt{1-\cos\alpha}}\sin\alpha\,\mathrm{d}\alpha \tag{B2}$$

with

$$\rho(\alpha) \; := \; \int\limits_0^\infty \frac{(\Psi\Psi*)(q,q,\alpha)}{q\sqrt{2}}\,\mathrm{d}q \qquad\qquad\qquad (B3)$$

The maximum entropy formalism now provides the probability distribution $\rho_{ME}$ which given the finiteness of the expectation on the right-hand side of eq. (B2) has maximal entropy corresponding to minimal angular correltation of teh electrons. The mathematics of the maximum entropy formalism [21] now gives

$$\rho_{ME} \; = \; e^{-\mu/\sqrt{1-\cos\alpha}}/N \qquad\qquad\qquad (B4)$$

where $\mu$ is a yet undetermined Lagrangian multiplier arising from the maximization of entropy under the constraint of the finiteness of the right-hand side of eq. (B2). It gets its final value of 0.6 by equating $\rho_{ME}(\pi)$ with $\rho_{00}(\pi) = 0.659$, i.e. through comparison of the maximum entropy distribution with the existing facts. The value of the normalization constant is

$$N \; = \; \int\limits_0^\pi e^{-0.6/\sqrt{1-\cos\alpha}}\sin\alpha\,\mathrm{d}\alpha \; = \; 0.992 \qquad\qquad\qquad (B5)$$

---